\documentclass[twocolumn,10pt]{IEEEtran}
\usepackage[utf8]{inputenc}
\usepackage{graphicx, flushend}
\usepackage{amsmath}
\usepackage{cases}
\usepackage{amssymb}
\usepackage{array}
\usepackage{subcaption}
\usepackage{array}
\usepackage{lipsum,color}
 \usepackage{epstopdf}

\DeclareGraphicsExtensions{.pdf,.png,.jpg,.eps,.ps}

\usepackage{algorithm}
\usepackage{algorithmicx}
\usepackage{algpseudocode}

\usepackage{amsthm,amssymb}

\usepackage{textcase}
\begin{document}

\pagestyle{empty}

\title{Joint  Computation Offloading and Resource Allocation in Cloud Based Wireless HetNets}

\author{\IEEEauthorblockN{Nguyen Ti Ti and Long Bao Le}
\vspace{-0.45cm}
\thanks{T. T. Nguyen and L.B. Le are with INRS-EMT, University of Qu\'{e}bec,  Montr\'{e}al, Qu\'{e}bec, Canada. Emails: \{titi.nguyen,long.le\}@emt.inrs.ca. }
}
%\author{\IEEEauthorblockN{Nguyen Ti Ti and Long Bao Le }
%	\IEEEauthorblockA{INRS-EMT, University of Quebec\\
%		 titi.nguyen@emt.inrs.ca, long.le@emt.inrs.ca}
%}

\maketitle

\thispagestyle{empty}
\begin{abstract}
In this paper, we study the joint computation offloading and resource allocation problem in the two-tier wireless heterogeneous network (HetNet). Our design aims to optimize the computation offloading to the cloud jointly with the subchannel allocation to minimize the maximum (min-max) weighted energy consumption subject to practical constraints on bandwidth, computing resource and allowable latency for the multi-user multi-task computation system. 
To tackle this non-convex mixed integer non-linear problem (MINLP), we employ the bisection search method to solve it
where we propose a novel approach to transform and verify the feasibility of the underlying problem in each iteration. 
%Moreover, we propose  to transform this problem into an integer linear program (ILP) which can be solved efficiently to achieve the global optimal solution. 
In addition, we propose 
a low-complexity algorithm, which can decrease the number of binary optimization variables and enable more scalable computation offloading
optimization in the practical wireless HetNets. 
Numerical studies confirm that the proposed design achieves the energy saving gains about 55\%
in comparison with the local computation scheme under the strict required latency of 0.1s. 
\end{abstract}


\IEEEpeerreviewmaketitle
\section{introduction}
The number of global devices has increased drastically in recent years. Moreover, with the multi-task processing capacity,
advanced mobile devices have led to proliferation of many computation-intensive applications covering different areas including entertainment, 
communication, social networking, e-health, image recognition, language processing, and gaming. 
These computation-extensive applications have demanded more powerful central processing unit (CPU) with higher clock
frequency, which will result in significant increase in the mobile energy consumption \cite{Kwak-2015}. 
Moreover, advancement in mobile battery technology is usually not sufficiently fast to
keep up with practical applications' requirements; therefore, the battery can become the bottleneck to improve the quality of experience (QoE)
for mobile users. Consequently, reducing mobile energy consumption in power-hungry applications is of great importance and 
one very potential solution for this problem is to offload heavy computations tasks to the edge cloud servers using the so-called
 mobile cloud computing (MCC) technology.
  
\vspace{0.2 cm} 
Recent development of small-cell based wireless HetNets promises enormous benefits from both the network and mobile user perspectives. 
First, deployment of low-power small cells enables efficient reuse of the radio spectrum, which helps increase 
the spectral efficiency. Second, the close transmitter-receiver proximity allows small cell users to
 achieve high signal-to-noise ratio (SNR) even with low transmit power. This enables them to 
meet the low-latency requirements of many emerging applications. Finally, realization of the MCC in small-cell based
wireless HetNets can lead to significant benefits such as prolonging battery lifetime and providing 
high-speed and ultra-low latency communications services in future 5G wireless systems.

Several MCC platforms have been proposed and developed in the literature such as MAUI \cite{MAUI}, CloneCloud \cite{Clone}, ThinkAir \cite{Think}
and a good survey for them with the corresponding computation offloading designs can be found in \cite{Barbarossa-2014}. In particular,  the tradeoff between transmission and computation energy was studied in  \cite{ti2017}, \cite{YouHCK16}. The joint computation task offloading scheduling and transmit power allocation of a single-user system was investigated in \cite{Mao0L17}.
Moreover, the authors in \cite{XuSCZ17} studied the multi-user radio resource management problem for the HetNet-MCC system,
which always offloads the entire computation task to the cloud. Dynamic computation offloading policies based on Lyapunov optimization 
were developed in \cite{Kwak-2015}, \cite{Don12}.  
These existing works, however, only consider the single-cell setting and many practical design aspects of the multi-cell MCC system such as 
dynamic computation offloading, joint multi-user resource allocation and computing resource assignment, and consideration of practical constraints
on bandwidth, operating frequency and tolerable delay limits are not satisfactorily accounted for. Our current work aims to fill this gap in the literature.

%Motivated by MCC and HetNet, we propose a MCC-HetNet system to enhance the energy efficiency of the mobile users. 

In this paper, we study the joint optimization problem for computation offloading and resource allocation where
computation tasks are either processed locally at the mobile or offloaded and processed in the cloud.
Moreover, offloaded tasks require radio resources for transmissions of the involved data (i.e., programming states). 
Importantly, this design is conducted for the multi-task multi-user multi-cell setting, which has not been
addressed in the literature to the best of our knowledge. The underlying joint computation task, subchannel, and operating clock speed
assignment problem, which aims to minimum the maximum weighted consumed energy subject to delay and resource constraints
is a non-convex and difficult MINLP problem. Therefore, we employ the bisection search method to solve it where
we transform the underlying non-convex INLP problem into an ILP for feasibility verification 
in each iteration. We also propose a low-complexity algorithm which is based on the decoupled optimization of
the resource assignments for the macrocell and small cells.

%This problem is converted to  via the proposed algorithms, and apply CVX solver to find the optimal solution. 

The remaining of this paper is organized as follows. Section II presents the system model, computation and transmission energy models
and problem formulation. Section III describes the solution method to solve the studied problem. Section IV 
evaluates the performance of proposed algorithms. Finally, Section V concludes the work.

\vspace{-0.1cm}
\section{System Model and Problem Formulation}

We consider a two-tier wireless HetNet with $M$ small cells (SC), one macro cell (MC), and $K$ users (UE). We assume that each cell is serviced by one base station (BS) connected to a common cloud provider via a high-speed fiber cable. Moreover, the cloud is assumed to have unlimited  computing resource to serve offloaded computation demands from multiple SUs simultaneously. We denote the set of all BSs as $\mathcal{M}_0 = \{0,\mathcal{M}\}$, where $0$ denotes the MC BS, and $\mathcal{M} =\{1,...,M\}$ denotes the set of SC BSs, and the set of all users is denoted as $\mathcal{K} = \{\mathcal{K}_0,\mathcal{K}_1,...,\mathcal{K}_M\}$, where $\mathcal{K}_0$ is the set of macro users (MUE)  and $\mathcal{K}_m$ is the set of SC users (SUE) associated with BS $m \in \mathcal{M}$. 

\vspace{-0.3cm}
\subsection{Computation Offloading Model}

We assume that UE $k_m \in \mathcal{K}_m, \; m\in \mathcal{M}_0$ has the set of $\mathcal{L}_{k_m}=\{1,2,...,l_{k_m}\}$ independent tasks
for his/her application and these tasks can be executed locally at the mobile device or offloaded and executed in the cloud independently 
over the time interval $T$ where $T$ is chosen to meet the delay requirement of the underlying application. 
Moreover, each task $l \in \mathcal{L}_{k_m}$ has the corresponding number of required CPU cycles $c_{k_m,l}$ (CPUs) and the number of
transmission bits of $b_{k_m,l}$ (bits) (i.e., to transmit the involved programming states). We now introduce a binary offloading 
decision variable for each task $l \in \mathcal{L}_{k_m}$ as $x_{k_m,l}$, which can be defined as follows:
\begin{eqnarray}
x_{k_m,l} = \begin{cases}
1, & \text{if task $l$ is executed at mobile device}\\
0, & \text{if task $l$ is offloaded to the cloud}
\end{cases}.
\end{eqnarray}

It is further assumed that the processor of UE $k_m \in \mathcal{K}_m$ has the DVFS (Dynamic Voltage and Frequency Scaling) capability
so it can adjust the operating frequency clock speed (FCS) in the clock speed set $f_{k_m} \in \mathcal{F}_{k_m}=\{f_{1}^{k_m},f_{2}^{k_m},...,f_{V_{k_m}}^{k_m}\}$ (CPU/s) 
which corresponds to the underlying application requirements \cite{Kwak-2015}. By introducing $V_{k_m}$  binary variables $\mu_v^{k_m}$, the operating 
clock speed can be expressed as $f_{k_m} = \sum_{v=1}^{V_{k_m}}\mu_v^{k_m}f_v^{k_m}$, where $\sum_{v=1}^{V_{k_m}}\mu_v^{k_m}=1$. Then, the local computation energy in Joule at UE $k_m \in \mathcal{K}_m$ can be expressed as
\begin{eqnarray}
E_c(k_m) = \frac{(\beta_{k_m,1}(f_{k_m} )^{\beta_{k_m,2}}+\beta_{k_m,3})\underset{l \in \mathcal{L}_{k_m}}{\sum}x_{k_m,l}c_{k_m,l}}{f_{k_m}},
\label{Ec}
\end{eqnarray}
where $\beta_{k_m,1},\beta_{k_m,2}, \beta_{k_m,3}$ denote the coefficients specified in the CPU model \cite{Kwak-2015}. 
For mobile devices such as Samsung Galaxy Note or Nexus, the value of $\beta_{k,2}$ is in range of [2.7-3]. In order to satisfy the application QoS, UE $k_m$ should complete its program execution within the delay  $\tau_{1,k_m} \leq T$ if such program is executed locally at the mobile device. We assume that the cloud has the replicated version of the execution file of the offloading tasks, which can be, therefore, executed in the cloud in the offloading case.  

\vspace{-0.25cm}
\subsection{Transmission Model}

We assume that the available spectrum is partitioned into non-overlapping bands, which are then allocated to the MC and SC tiers to
avoid the cross-tier interference. Moreover, the spectrum allocated to small cells is assumed to be fully reused over these cells (i.e., the SC
spectrum reuse factor is one). Moreover, the OFDMA scheme is assumed where the set of available sub-channels is denoted as $\mathcal{N} = \{1,2,...,N\}$.

Let $h_{k_j,m}^{(n)}$ denote the channel gain from UE $k_j \in \mathcal{K}_j$ to  BS $m \in \mathcal{M}_0$ on subchannel $n \in \mathcal{N}$. The transmit power of user $k_j \in \mathcal{K}$ on each subchannel is assumed to be the constant $P_t W$, where $W$ is the bandwidth of each subchannel, and the noise power density on subchannel $n$  at BS $m$ is $\sigma_{m}^{(n)}$.
We represent the subchannel assignment by a binary variable $\rho_{k_m}^{(n)}$, where $\rho_{k_m}^{(n)} = 1$ if subchannel $n$ is assigned to UE $k_m \in \mathcal{K}_m$, and $\rho_{k_m}^{(n)} = 0$, otherwise. Then, the transmission rate from UE $k_m$ to the BS $m$ can be expressed as
\begin{eqnarray}
r_{k_m} =  \underset{n \in \mathcal{N}}{\sum}\rho_{k_m}^{(n)}W\log _2\big(1+\gamma_{k_m}^{(n)}\big),
\label{rkm}
\end{eqnarray}
where
$
\gamma_{k_m}^{(n)} = \begin{cases} \frac{P_th_{k_m,m}^{(n)}}{\sigma_{m}^{(n)}}, & \text{if} \; m=0\\
\frac{P_th_{k_m,m}^{(n)}}{\sum_{j \neq m}\sum_{k_j \in \mathcal{K}_j}\rho_{k_j}^{(n)}P_th_{k_j,m}^{(n)} +\sigma_{m}^{(n)}}, & \text{if} \; m\neq 0
\end{cases}.
$

\vspace{0.3cm}
The power $P$ required to transmit data related to the program states from each mobile user comprises two parts, namely circuit power  $P_c$ (W/Hz) and
transmit power $P_t$ (W/Hz), which can be expressed as $P = P_c+P_t $.  Then, the total required energy related to the  transmission of
 UE  $k_m \in \mathcal{K}_m$  can be computed as follows:
\begin{eqnarray}
	E_t(k_m) = t_{k_m}PW
\sum\limits_{n \in \mathcal{N}}\rho_{k_m}^{(n)},
\label{Et}
\end{eqnarray}
where $t_{k_m}$ is the transmission time of the program states from the mobile to its BS, which can be calculated as
\begin{eqnarray}
t_{k_m}=\frac{1}{r_{k_m}} \sum\limits_{l \in \mathcal{L}_{k_m}} (1-x_{k_m,l})b_{k_m,l}.
\label{tkm}
\end{eqnarray}
The total latency experienced by an offloaded task comprises the time required for sending program states/bits to the cloud, the computation time in
the cloud, and the time required for downloading the results to the mobile. However, cloud computation time is relatively small
due to the high cloud computation power and the data related to computation results' download has much smaller size compared to
 the offloading data in general. Therefore, we neglect the cloud energy consumption and data download transferred time. 
Moreover, to ensure the constrained latency, the transmit time $t_{k_m}$ is
required to be smaller than the maximum delay, i.e., $\tau_{2,k_m} < T$.

\vspace{-0.15cm}
\subsection{Problem Formulation}

We now present the formulation for the considered problem where our design objective is to minimize the maximum weighted users' consumed energy.
The energy weight, denoted as $w_{k_m}$, represents the priorities or the battery/computation levels of different users \cite{Guo1}.
Then, the joint computation offloading and resource allocation problem with latency, radio and computational resource constraints can be 
stated as

\begin{eqnarray} 
\begin{aligned}
(\mathcal{P}_1)  \; \; & \underset{\boldsymbol{\rho,x,\mu}}{\min}  \; \underset{k_m} {\max} \; w_{k_m}(E_c(k_m) + E_t(k_m)) \nonumber \\
 \text{subject to} &\\
 (\text{C}1):& \;  t_{k_m} \leq \tau_{2,k_m},  \; \forall m \in \mathcal{M}_0, \forall k_m \in \mathcal{K}_m \\
(\text{C}2):& \; \sum_{k_m \in \mathcal{K}_m} \rho_{k_m}^{(n)} + \sum_{k_0 \in \mathcal{K}_0}\rho_{k_0}^{(n)} \leq 1, \forall m \neq 0, \forall n \in \mathcal{N}\\
  \end{aligned}	
\end{eqnarray}
\begin{eqnarray} 
\begin{aligned}
(\text{C}3):& \; \rho_{k_m}^{(n)} \in \{0,1\} , \forall m \in \mathcal{M}_0, \forall k_m \in \mathcal{K}_m, \forall n \in \mathcal{N}\\
(\text{C}4):& \; \frac{\underset{l \in \mathcal{L}_{k_m}}{\sum}x_{k_m,l}c_{k_m,l}}{f_{k_m}} \leq \tau_{1,k_m}, \; \forall m 
\in \mathcal{M}_0, \forall k_m \in \mathcal{K}_m \nonumber \\
(\text{C}5):& \; x_{k_m,l} \in \{0,1\},\; \forall m \in \mathcal{M}_0, \forall k_m \in \mathcal{K}_m , \forall l \in \mathcal{L}_{k_m} \nonumber \\
(\text{C}6):& \; f_{k_m} = \sum_{v=1}^{V_{k_m}}\mu_v^{k_m}f_v^{k_m},\; \mu_v^{k_m} \in \{0,1\}, \forall k_m \in \mathcal{K}_m \\
(\text{C}7):& \; \sum_{v=1}^{V_{k_m}}\mu_v^{k_m}=1 , \forall k_m \in \mathcal{K}_m .
\end{aligned}	
\end{eqnarray}

In this problem formulation, constraint (C1)  captures the transmission latency requirements for offloading process. Constraint (C2) and (C3) represent the MC and SCs bandwidth sharing where
each subchannel can be allocated to at most one MUE or one SUE in each SC. Constraint (C4) represents the delay requirements for local computation.
Furthermore, constraint (C5) captures the binary offloading decisions while the remaining constraints express the computational capacity of mobile devices.

\section{Algorithm Development}\label{algsect}

The considered problem $(\mathcal{P}_1)$ is indeed a non-convex INLP  due to the integer optimization variables for allocating tasks, frequency clock speeds' 
selection and subchannel assignments and due to the non-convexity of the objective function and constraint functions in (C1).
Therefore, this problem is very difficult to solve. To have an insightful description, we first reformulate the min-max objective of $(\mathcal{P}_1)$ as follows:
\begin{eqnarray}
\begin{aligned}
(\mathcal{P}_2)  \; \; &\min \zeta \nonumber \\
\text{subject to} &\\
(\text{C}8): & \;  w_{k_m}(E_c(k_m) + E_t(k_m)) \leq \zeta ,\; \forall
k_m\\
(\text{C}1)-&(\text{C}7).
\end{aligned}	
\label{prob2}
\end{eqnarray}

\subsection{Proposed Algorithm (Optimal alg.)}

We can now apply the bisection search method to find the optimal min-max users' energy consumption for the reformulated problem $(\mathcal{P}_2)$. 
Specifically, the bisection search method iteratively updates an upper-bound $\zeta_{\max}$ and a lower-bound $\zeta_{\min}$ of the objective value $\zeta$ of problem $(\mathcal{P}_2)$. In particular, in each iteration, we have to verify the feasibility of 
problem $(\mathcal{P}_2)$ for a given value of $\zeta$ based on which we can update $\zeta_{\max}$ and $\zeta_{\min}$.
 If the set of constraints is feasible, then upper-bound of objective function will decrease, and inversely its lower-bound will increase. This algorithm will terminate when the difference between upper-bound and lower-bound values becomes sufficiently small. The proposed algorithm which can find the optimal solution of $(\mathcal{P}_2)$ is given in Algorithm 1.

In order to verify the feasibility of problem $(\mathcal{P}_2)$, we take three major steps 
to transform all constraints of problem $(\mathcal{P}_2)$ into the linear form. In the first step,
we linearize the involved logarithmic functions in (C1) and (C8). In the second step, we attempt
to determine whether UEs can locally process their tasks or not for a given value of $\zeta$. 
% transform the fractional constraint functions into the non-fractional form by eliminating the denominators of  
%certain terms in these functions. In order to do so, we determine whether these involved denominators
%are zero or not by solving an auxiliary problem. 
In the final step, we introduce some further auxiliary variables
to transform the product-form of the obtained constraint functions into the desirable linear form. The obtained
linear program after step three can then be solved effectively.

These steps are described in more details for a given value of $\zeta$ in the following.

\subsubsection{Step one} 

We introduce some auxiliary binary variables as follows:
\begin{eqnarray}
\alpha_{k_1,k_2,...,k_M}^{(n)} = \begin{cases}
1 , & \text{if} \; \prod_{m \in \mathcal{M}}\rho_{k_m}^{(n)} = 1, \; k_m \in \mathcal{K}_m\\
0, & \text{otherwise}
\end{cases}.
\label{alpha}
\end{eqnarray}
The above expression means that the variable $\alpha_{k_1,k_2,...,k_M}^{(n)}$ will be active if users $k_1 \in \mathcal{K}_1, k_2 \in \mathcal{K}_2,...,k_M \in \mathcal{K}_M$ transmit on the same subchannel $n$. We have to now re-write $E_t({k_m})$ in (\ref{Et}), which depends on $t_{k_m}$ given in (\ref{tkm}).
Toward this end, the transmission rate from SUE $k_m \in \mathcal{K}_m$ to BS $m \in \mathcal{M}$ is re-expressed in (\ref{rkmm})
which is needed in the expression of $t_{k_m}$. We also need (\ref{rho4}) to re-write $E_t({k_m})$ in (\ref{Et}). Moreover, 
constraint (C$2$) can be now rewritten as in (\ref{rho3}). 

\newcounter{tempequationcounter_1}
\begin{figure*}[!t]
	\normalsize
	\setcounter{tempequationcounter_1}{\value{equation}}
	\begin{IEEEeqnarray}{rCl}
		r_{k_m} = \underset{n \in \mathcal{N}}{\sum}\underset{k_1 \in \mathcal{K}_1}{\sum}...\underset{k_{m-1} \in \mathcal{K}_{m-1}}{\sum}\underset{k_{m+1} \in \mathcal{K}_{m+1}}{\sum}...\underset{k_M \in \mathcal{K}_M}{\sum}&\alpha_{k_1,k_2,...,k_M}^{(n)}&W\log_2(1+\frac{P_th_{k_m,m}^{(n)}}{\sum_{j \in \mathcal{M}\setminus m}P_th_{k_j,m}^{(n)}+\sigma_{m}^{(n)}}) .		\label{rkmm}\\
		\underset{k_1 \in \mathcal{K}_1}{\sum}...\underset{k_M \in \mathcal{K}_M}{\sum}\alpha_{k_1,k_2,...,k_M}^{(n)} + \underset{k_0 \in \mathcal{K}_0}{\sum}\rho_{k_0}^{(n)} &\leq & 1,\quad \forall n \in \mathcal{N}. \label{rho3} \\
		\underset{n \in \mathcal{N}}{\sum}\rho_{k_m}^{(n)} = \underset{n \in \mathcal{N}}{\sum}\underset{k_1 \in \mathcal{K}_1}{\sum}...\underset{k_{m-1} \in \mathcal{K}_{m-1}}{\sum}\underset{k_{m+1} \in \mathcal{K}_{m+1}}{\sum}&...&\underset{k_M \in \mathcal{K}_M}{\sum}\alpha_{k_1,k_2,...,k_M}^{(n)}. \label{rho4} \\
		r_{k_m}\sum_{v=1}^{V_{k_m}}\sum_{l \in \mathcal{L}_{k_m}}\mu_{v}^{k_m}F_v^{k_m}x_{k_m,l}c_{k_m,l} + \sum_{l \in \mathcal{L}_{k_m}}(1-x_{k_m,l})&b_{k_m,l}PW&\sum_{n \in \mathcal{N}}\rho_{k_m}^{(n)} \leq \frac{\zeta r_{k_m}}{w_{k_m}}, \text{if} \; r_{k_m}>0, \; \forall k_m . \label{reE}
	\end{IEEEeqnarray}
	%\addtocounter{tempequationcounter}{2}
	\setcounter{equation}{\value{tempequationcounter_1}}
	\hrulefill
	\vspace*{1pt}
\end{figure*}
\addtocounter{equation}{4}
\begin{algorithm}[t]
	\caption{Multi-task and Multi-user Computation Offloading and Resource Allocation }
	\begin{algorithmic}[1]
		\State \textbf{Initialize}: choose $\epsilon$, $\zeta_{\min} = 0$ and $\zeta_{\max}= \underset{m \in \mathcal{M}, k_m \in \mathcal{K}_m}{\max}E_c(k_m)|\{x_{k_m,l}=1, \forall l \in \mathcal{L}_{k_m}\}$.
		\While{$\zeta_{\max}-\zeta_{\min}<\epsilon$}
		\State Compute $\zeta = (\zeta_{\max}+\zeta_{\min})/2$.
		\For{each user $k_m$} 
			\If{optimal value of $\mathcal{P}_{s1}^{k_m}=0$}
			\State neglect user $k_m$.
			\Else
			\State Assign $x_{k_m,l} = 0$ as in \textit{Proposition 2}.
			\EndIf
		\EndFor
		%\If{LC1}
		%\State Solve problem $\mathcal{P}_{s2}$ of MC to obtain the set $\mathcal{N}_{SC}$.
		%\State Check the feasibility of $(\mathcal{P}_2)$  with only SC users with the set of available subchannels $\mathcal{N}_{SC}$ as in Section \ref{algsect}.3.
		%\ElsIf{LC2}
		%\State Check the feasibility of $(\mathcal{P}_2)$ for all users when SINR is given in (\ref{gamma1}).
		%\Else
		\State Check the feasibility of $(\mathcal{P}_2)$  for users (optimal value of $\mathcal{P}_{s1}^{k_m} \neq 0, \forall k_m$)  with the set of available subchannels $\mathcal{N}$ as in Section \ref{algsect}.A.3.
		%\EndIf
		\If{feasibility}
		\State Assign $\zeta_{\max} = \zeta$.
		\Else
		\State Assign $\zeta_{\min} = \zeta$.
		\EndIf
			
		\EndWhile\label{alg2}
		
		%\If {LC2}
		%\State Compute $\alpha_{k_1,k2,...,k_M}^{(n)}, \forall k_m \in \mathcal{K}_m$ as in (\ref{alpha}) and update transmission rate $r_{k_m}$ as in (\ref{rkmm}) for SUEs.
		%\State Reallocate the computing resource for each SUE with the updated value of transmission rate by the same transformation as in Section \ref{algsect}.1.
		%\EndIf
	\end{algorithmic}
\end{algorithm}

\subsubsection{Step two} For a given $\zeta$, UEs will not offload their tasks if they can process all tasks locally. Therefore, to determine whether UEs offload or not, we find the minimum number of transmission bits of UE $k_m$ as $\sum_{l \in \mathcal{L}_K}(1-x_{k_m,l})b_{k_m,l}$ when its computing energy is less than $\zeta$.  If this value is equal to zero, UE $k_m$ can locally execute its application; therefore the transmission rate $r_{k_m}$ will be qual to zero. This problem is formulated as follows:

%In order to transform the fractional form due to $E_t(k_m)$ in constraints (C8) to the linear form,
%we first re-write $E_t(k_m)$ as follows:
%
%\begin{eqnarray}
%	E_t(k_m) =P W \frac{\underset{l \in \mathcal{L}_{k_m}}{\sum}(1-x_{k_m,l})b_{k_m,l}}{r_{k_m}}  \underset{{n \in \mathcal{N}}}{\sum}\rho_{k_m}^{(n)}.
%\label{Et1}
%\end{eqnarray}

%where $t_{k_m}$ is the transmission time of the program states from the mobile to its BS, which can be calculated as
%\begin{eqnarray}
%t_{k_m}= \frac{\underset{l \in \mathcal{L}_{k_m}}{\sum}(1-x_{k_m,l})b_{k_m,l}}{r_{k_m}}.
%\label{tkm}
%\end{eqnarray}

%We can see that $r_{k_m}$ is in the denominator of this expression of $E_t(k_m)$ and $r_{k_m}$ will be equal to zero
%if the required tasks of user $k_m$ are executed locally by the mobile user whereas $r_{k_m}>0$ if at least one task
%is offloaded to the cloud. Because of this, we determine if the tasks of each user $k_m$ are executed locally or not.
%This can be fulfilled by solving the following problem, which minimize the number of transmission bits $\sum_{l \in \mathcal{L}_K}(1-x_{k_m,l})b_{k_m,l}$ for the task offloading case. 
\begin{eqnarray}
\begin{aligned}
(\mathcal{P}_{s_1}^{k_m})  \; \; &\min\limits_{\boldsymbol{x}_{k_m},\boldsymbol{\mu}_{k_m}} \sum_{l \in \mathcal{L}_{k_m}}(1-x_{k_m,l})b_{k_m,l} \nonumber \\
\text{subject} & \text{ to}\\
(\text{C}9):  \;  &w_{k_m} E_c(k_m)  \leq \zeta,\\
(\text{C}4) -& (\text{C}7) .
\end{aligned}	
\label{prob3}
\end{eqnarray}
%The constraint (C9) requires that the locally computational energy does not exceed $\zeta_1$.

%\subsubsection{Optimal Solution for Problem $\mathcal{P}_{s_1}^{k_m}$}

To solve problem $\mathcal{P}_{s_1}^{k_m}$, we re-express one term in computation energy expression as follows:

\begin{eqnarray}
\frac{(\beta_{k_m,1}(f_{k_m} )^{\beta_{k_m,2}}+\beta_{k_m,3})}{f_{k_m}} = \sum_{v=1}^{V_{k_m}}\mu_v^{k_m}F_v^{k_m},
\label{Fv}
\end{eqnarray}
where $F_v^{k_m} = \frac{(\beta_{k_m,1}(f_v^{k_m} )^{\beta_{k_m,2}}+\beta_{k_m,3})}{f_v^{k_m}}$ for $f_v^{k_m} >0$ and $F_v^{k_m} = 0$  for $f_v^{k_m}=0$. The constraint (C$9$) now is the sum of the 
product of two binary variables, which can be given as
\begin{eqnarray}
(\sum_{v=1}^{V_{k_m}}\mu_v^{k_m}F_v^{k_m})(\underset{l \in \mathcal{L}_{k_m}}{\sum}x_{k_m,l}c_{k_m,,l}) \leq \frac{\zeta}{w_{k_m}}.
\label{eq16}
\end{eqnarray}
We now deal with the non-convex term $z_{k_m,v,l}=\mu_v^{k_m}x_{k_m,l}$ in (\ref{eq16}). In general, the product of binaries variables can be converted to the linear inequalities as suggested in \cite{LeeMINLP}. Particularly, the $0/1$-variable $y = \prod\limits_{i=1}^{n}s_i$ can be expressed equivalently as
\begin{eqnarray}
\begin{cases}
y &\in \{0,1\} ,\; s_i \in \{0,1\}, \forall i \\
y  &\geq \sum\limits_{i=1}^n s_i -n + 1\\
y &\leq \min \{s_i\} \\ 
\end{cases}.
\label{qkm1}
\end{eqnarray}
Then applying (\ref{qkm1}), we can transform the non-convex term $z_{k_m,v,l}=\mu_v^{k_m}x_{k_m,l}$ to linear form of $z_{k_m,v,l},\mu_v^{k_m}$ and $x_{k_m,l}$.
%\begin{eqnarray}
%\begin{cases}
%z_{k_m,v,l} &\in \{0,1\} \\
%z_{k_m,v,l} &\geq \mu_v^{k_m}+x_{k_m,l}-1\\
%z_{k_m,v,l} &\leq \mu_v^{k_m}  \\
%z_{k_m,v,l} &\leq x_{k_m,l}
%\end{cases}.
%\label{zkm}
%\end{eqnarray}
In addition, constraints (C$4$) can be easily converted
 to a linear form as: $
\underset{l \in \mathcal{L}_{k_m}}{\sum}x_{k_m,l}c_{k_m,,l}- \tau_{1,k_m}\sum_{v=1}^{V_{k_m}}\mu_v^{k_m}f_v^{k_m} \leq 0.$

Using these expressions, the considered problem can be transformed into an ILP with optimization variables $\boldsymbol{x,z,\mu}$ which can be solved
 effectively by using the interior-point method or the solver CVX-Gurobi \cite{LeeMINLP}. 

\subsubsection{Step three - feasibility verification for problem $\mathcal{P}_{2}$}
We now state some important results in the following two propositions, which correspond to two cases where
the offloading decision  variables $x_{k_m,l}$ are zero and one, respectively.

\textit{\textbf{Proposition 1}}: If there exists a feasible solution for $\mathcal{P}_2$  given $\zeta$
 and the optimal value of $\mathcal{P}_{s_1}^{k_m}$ 
 %(denoted as $d_{k_m}^*$)
  is 0, then the offloading decision
 variables $x_{k_m,l}, \forall l \in \mathcal{L}_{k_m}$ are set equal to 1 and this will form a feasible solution.

\begin{proof} It is clear that the efficient optimization for  $x_{k_m},\mu_{k_m}$ in $\mathcal{P}_2$ must allocate the smallest
amount of radio resources to meet the fixed energy level $\zeta$. In fact, if any user, who has
 computing energy less than $\zeta$ and has execution time satisfying the computing delay time, offloads its tasks to the cloud, 
it will occupy the radio resources of other users demanding for computation offloading, which results in the increase of transmit energy of those users. This proves the proposition. \end{proof}

\textit{\textbf{Proposition 2}}: If there exists a feasible solution for $\mathcal{P}_2$ given $\zeta$ and the optimal value of $\mathcal{P}_{s_1}^{k_m}$ is positive, then the offloading decision variable $x_{k_m,l}$ is set equal to 0 if the optimal solution $x_{k_m,l}$ of $\mathcal{P}_{s_1}^{k_m}$ equals to 0
and this will form a feasible solution.

\begin{proof} If the optimal value of $\mathcal{P}_{s_1}^{k_m}$ is greater than 0, it means that the local computational energy must be less than $\frac{\zeta}{w_{k_m}}$ since the total computation and transmission energy must be less than or equal to $\frac{\zeta}{w_{k_m}}$. Because the objective of $\mathcal{P}_{s_1}^{k_m}$ is to minimize the number of transmission bits, UE $k_m$ will consume the least transmission energy with a given radio resource. Therefore, the task offloading
decision variables must be set equal to 0 (offload to cloud) if the solution of task allocation of $\mathcal{P}_{s_1}^{k_m}$ is equal to 0. Note that when the optimal value of  $\mathcal{P}_{s_1}^{k_m}$ is positive, the transmission rate $r_{k_m}$ must be greater than 0 to offload data to the cloud. Therefore, we can rewrite the fractional constraint functions of (C$8$) into non-fractional form as (\ref{reE}). \end{proof}

Using the results in \textit{Proposition 1}, we can set the zero rate for users with  $x_{k_m,l}$ equal to one for all tasks $l$.
For remaining users whose optimal $x_{k_m,l}$ are equal zero for at least one task according to \textit{Proposition 2}, 
we solve problem ${\mathcal{P}_{s2}}$ to determine the computation offloading and resource allocation solution. 

The remaining thing is to verify the feasibility of the resulting equivalent problem of the original  problem  $\mathcal{P}_2$. 
Applying the same technique as in (\ref{qkm1}) for UEs having transmission rate $r_{k_m} > 0$, we transform the product-form of (\ref{reE}) into linear-form of  $x_{k_m,l}$,  $\mu_v^{k_m}$, $\alpha_{k0,0}^{(n)}$,  $q_{k_0,v,l}^{(n)}$, $u_{k_0,l}^{(n)}$,  $\alpha_{k_1,...k_M}^{(n)}$, $q_{k_1,k_2,...,k_M,v,l}^{(n)}$ and $u_{k_1,k_2,...,k_M,l}^{(n)}$, where $q_{k_1,k_2,...,k_M,v,l}^{(n)} = \mu_{v}^{k_m}x_{k_m,l}\alpha_{k_1,k_2,...,k_M}^{(n)}$, $u_{k_1,k_2,...,k_M,l}^{(n)}=x_{k_m,l}\alpha_{k_1,k_2,...,k_M}^{(n)}$ for SUEs,  $q_{k_0,v,l}^{(n)} = \mu_v^{k_0}x_{k_0,l}\rho_{k_0}^{(n)}$, and $u_{k_0,l}^{(n)}=x_{k_0,l}\rho_{k_0}^{(n)}$ for MUEs.

Using these transformations, the set of constraints of problem $(\mathcal{P}_2)$
can be converted to the linear form of  $x_{k_m,l}$,  $\mu_v^{k_m}$, $\alpha_{k0,0}^{(n)}$,  $q_{k_0,v,l}^{(n)}$, $u_{k_0,l}^{(n)}$,  $\alpha_{k_1,...k_M}^{(n)}$, $q_{k_1,k_2,...,k_M,v,l}^{(n)}$ and $u_{k_1,k_2,...,k_M,l}^{(n)}$. Therefore, the feasibility verification of the transformed problem can be done effectively by a standard solver.

%\subsection{Low-Complexity Algorithms}
\subsection{Low-complexity Algorithm with Decoupled MC-SC Optimization (LC alg.)}

We now present a low-complexity algorithm, which can perform well in large-scale wireless HetNets.
In this algorithm, we first determine the minimum number of subchannels so that all MC UEs (MUEs) can meet the energy consumption level $\zeta$,
 which can be stated as follows:
\begin{eqnarray}
\begin{aligned}
(\mathcal{P}_{s2})  \; \;\min  &\sum_{n \in \mathcal{N}}\sum_{k_0 \in \mathcal{K}_0}\rho_{k_0}^{(n)} \nonumber  \\
\text{subject to} &\\
(\text{C}1), (\text{C}3)-&(\text{C}8), \text{for} \; m=0.
\end{aligned}	
\label{prob4}
\end{eqnarray}

After solving problem $\mathcal{P}_{s2}$ by using the above transformations for MUEs, 
 the set of remaining subchannels that SC UEs (SUEs) can use can be written as $\mathcal{N}_{SC} = \mathcal{N}\setminus\{n| {\rho_{k_0}^{(n)}}_{(\mathcal{P}_{s2})} = 1, \forall k_0 \in \mathcal{K}_{0}\}$. 

We can then allocate these remaining subchannels ($n \in \mathcal{N}_{SC}$) by solving problem $\mathcal{P}_2$ with only SUEs $k_m \in \mathcal{K}_m, \forall m \in \mathcal{M}$ (i.e., we remove any terms related to MUEs in this problem). This problem can be solved by using the proposed Algorithm 1 for only SUEs with the available subchannels $\mathcal{N}_{SC}$.
The number of optimization variables in this algorithm now decreases $\frac{|\mathcal{N}|}{|\mathcal{N}_{SC|}}$ times in compassion with the
case where the joint subchannel allocation optimization for MC and SCs is conducted. 

\section{Numerical Results}
\vspace{-0.3cm}
\begin{figure} [!h]
	\centering	
	\includegraphics[width=0.3\textwidth]{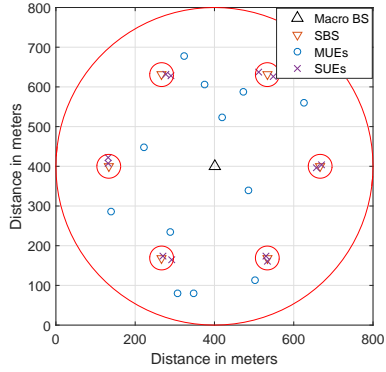}
	\caption{Network topology and user placement in the numerical examples.}
	\label{fig1}
\end{figure}
\vspace{-0.1cm}

\begin{figure} [!h]
	\centering	
	\includegraphics[width=0.37\textwidth]{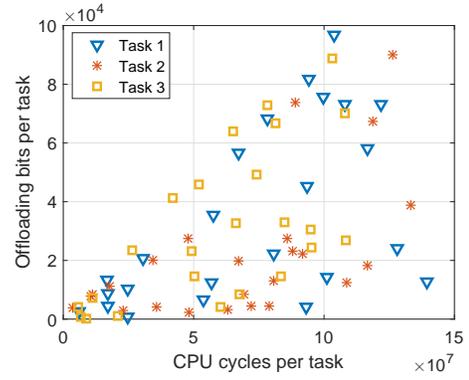}
	\caption{Computation - transmission relation of computation tasks.}
	\label{fig2}
\end{figure}

\begin{figure} [!h]
	\centering	
	\includegraphics[width=0.37\textwidth]{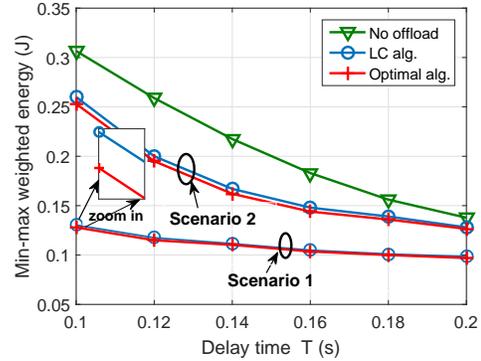}
	\caption{Min-max weighted energy consumption.}
	\label{fig3}
\end{figure}

The network setting used in our simulations is shown in Fig.\ref{fig1}, where there are 12 MUEs in macrocell and 2 SUEs for each small cell. The MC and SC 
coverage radius are $400$m and $30$m, respectively. All UEs have 11 levels of operating clock frequency uniformly chosen in $0-2$ GHz. For convenience, we assume that all users has 3 tasks and total CPU requirement for each user is $0.2$ Gcycles. The maximum tolerable computing delay is set equal to  $T$ for all users while the transmission delay is set randomly in $0.7T-0.9T$. The number of transmission bits/task and CPU cycles/task are illustrated in Fig.\ref{fig2},
which are used in scenario 1 (presented in Fig. \ref{fig3}) in which the ratio of  $b_{k_m,l}/c_{k_m,l}$ is chosen randomly in $10^{-5}-10^{-3}$ 
(as in Fig. \ref{fig2}). The energy weight is set randomly in $0.8-1$.  The energy coefficients are set for all users as $\beta_{k_m,1} = 0.34(10^{-9})^{\beta_{k_m,2}}, \; \beta_{k_m,2} = 3 $ and $\beta_{k_m,3} = 0.35$ \cite{Kwak-2015}.
\pagebreak

%\vspace{-0.4cm}
%\vspace{-0.2cm}

We set the number of subchannels as $20$, and the bandwidth per subchannel is $180$ kHz, and the noise power density equal to $-140$ dBm/Hz. The transmission 
power  $P_t$ is set equal to $-33$ dBm/Hz and  $-43$ dBm/Hz for MUEs and SUEs, respectively  and the circuit power $P_c = P_t/2$. The subchannel gains are generated according to $h_{k_m,m}^{(n)} = \xi^{(n)}g_{k_m,m}$ where $\xi^{(n)}$ is a random value generated according to the exponential distribution and $g_{k_m,m}$ denotes the pathloss defined according  3GPP technical report as $g_{k_0,0} = -128.1 - 37.6\log10(d_{k_0,0}) $ (dB) for MUEs and   $g_{k_m,m} = -127- 30\log10(d_{k_m,m}), \forall m \neq 0 $ (dB) for SUEs \cite{3GPP} where $d_{k_m,m}$ is the geographical distance between UE $k_m$ and BS $m$ (km). The stop condition of bisection search is set as $\epsilon = 10^{-3}$.
\vspace{-0.3cm}

Fig. \ref{fig3} shows the min-max weighted energy required to execute the applications of all UEs, which is obtained by averaging
the result over 15 system realizations under the proposed optimal (Optimal alg.) and low-complexity (LC alg.) and no computation offload (No offload). 
The ratio of bits per CPU cycle (BPC) in scenario 2 is 50 percent higher than that in scenario 1. 
These results show that the smaller the BPC, the smaller the consumed energy for all schemes. For the computation load of $0.2$ Gcycles per user, 
the min-max weighted energy without offloading is much higher than that due to the proposed schemes under both scenarios. In particular, the proposed
optimal algorithm can reduce the energy about $55\%$ compared with the ``No offload" scheme in scenario 1. Moreover, the energy consumption in 
the LC scheme is nearly equal to the global optimal solution due to the ``Optimal alg.".
\pagebreak
\begin{figure} [!h]
	\centering	
	\includegraphics[width=0.37\textwidth]{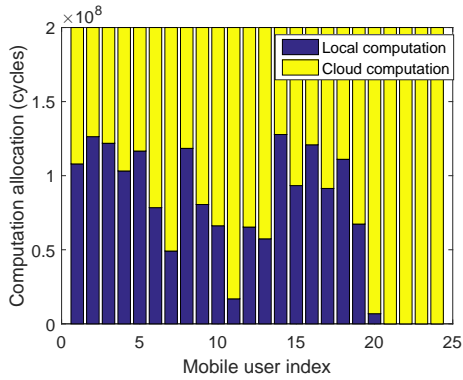}
	\caption{Computation allocation with latency of 0.1s.}
	\label{fig4}
\end{figure}
\vspace{-0.2cm}

The computation allocation using LC alg. for different users with  $T = 0.1$s in one system realization is illustrated 
in Fig. \ref{fig4}. This figure shows that the computation load is distributed fairly equal between users thanks to min-max weighted energy design objective. 
Moreover, some UEs, having  small BPC tasks or high SINR ratio, offload all tasks to the cloud. 
This figure also shows that the worst UE can offload 0.06 (Gcycles), then its FCS decreases about 30 percent leading
 to the decrease of  computation energy  by nearly 2.4 times.

\vspace{-0.36cm}
\begin{figure} [!h]
	\centering	
	\includegraphics[width=0.37\textwidth]{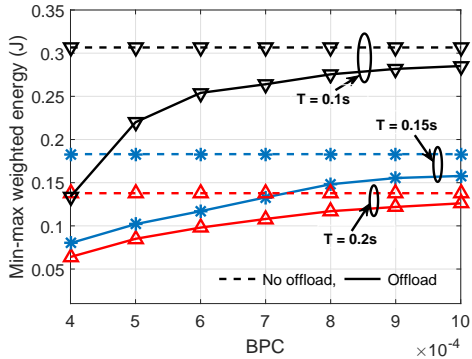}
	\caption{Min-max weighted energy versus bits per 1 CPU cycle.}
	\label{fig5}
\end{figure}
\vspace{-0.2cm}

Fig. \ref{fig5} shows the min-max weighted energy consumption when the number of cycles per task is fixed by $0.2/3$ (Gcycles) while the 
BPC is set the same for all UEs. When this parameter is small, the performance gap in terms of min-max energy consumptions 
between the proposed offloading (``offload") design and ``No offload" scheme is quite large. This means that the proposed scheme can result in great energy reduction. 

\vspace{-0.36cm}
\begin{figure} [!h]
	\centering	
	\includegraphics[width=0.37\textwidth]{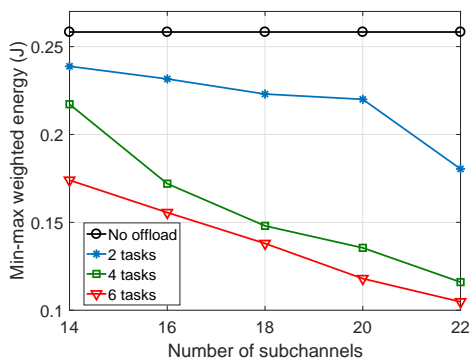}
	\caption{Min-max weighted energy consumption for 0.2 Gcycles computation load.}
	\label{fig6}
\end{figure}

\pagebreak
Fig. \ref{fig6} shows the achievable min-max weighted energy for scenarios where each user has  2, 4 and 6 tasks and the total computation load is 0.2 Gcycles
 per user, maximum tolerable  latency $T=0.12s$ while the BPC is fixed at $5\times10^{-4}$. It can be observed that  the  min-max weighted energy
 decreases quite drastically as the number of tasks increases. 
Moreover, if there are two tasks then users send at least one task to the cloud until $N=22$ subchannels  while if there are 4 tasks then users can send at least one task to the cloud when $N \geq 16$. 
 However, in all cases, if sufficient radio resources are available, the UE with largest weighted energy prefers to offload first to achieve the 
lowest min-max weighted energy. Therefore, the ``No offload" have the worst performance in term of energy consumption comparing with the proposed schemes.

\vspace{-0.3cm}
\section{Conclusion}

In this paper, we have proposed a general framework for multi-task multi-user multi-cell computation offloading. Considering
 the practical discontinuity of operating frequency clock speed of real-world chipsets, and the partitioning
 of computation load into individual tasks, we have formulated the problem which minimizes the maximum weighted energy consumption while maintaining the application latency requirement. We have then developed the optimal and low-complexity algorithms to tackle this problem. Numerical results have confirmed the desirable performance of the proposed algorithms for wireless HetNets which can lead to great saving of the energy consumption.
\vspace{-0.3cm}
\bibliographystyle{IEEEtran}  
\bibliography{MCCpaper}

\end{document}